\documentclass[10pt,prd,twocolumn,floatfix,preprintnumbers,nofootinbib,superscriptaddress,longbibliography]{revtex4-1}
\pdfoutput=1
\usepackage[colorlinks=true,breaklinks=true]{hyperref}
\hypersetup{allcolors=[rgb]{0.0 0.0 0.6},linkcolor=[rgb]{0.0 0.5 0.6}}
\usepackage[normalem]{ulem}
\usepackage{amsmath,amssymb}
\usepackage{epsfig}  
\usepackage{graphicx}   
\usepackage{slashed}             
\usepackage{url}
\usepackage{color}
\usepackage{multirow}
\usepackage{placeins}
\usepackage[dvipsnames]{xcolor}
\usepackage{letltxmacro}
\DeclareMathOperator{\GeV}{GeV}
\LetLtxMacro{\oldcite}{\cite}
\renewcommand{\cite}[1]{\mbox{\oldcite{#1}}}

\usepackage{comment}

\usepackage{lipsum, babel}

\usepackage{empheq,etoolbox}
\patchcmd{\subequations}
  {\theparentequation\alph{equation}}
  {\theparentequation.\alph{equation}}
  {}{}

\newcommand{\keV}{\text{ keV}}

\begin{document}

\title{Constraining heavy axion-like particles by  energy deposition  in Globular Cluster stars}

\author{Giuseppe Lucente}\email{giuseppe.lucente@ba.infn.it}
\affiliation{Dipartimento Interateneo di Fisica “Michelangelo Merlin,” Via Amendola 173, 70126 Bari, Italy}
\affiliation{Istituto Nazionale di Fisica Nucleare - Sezione di Bari, Via Orabona 4, 70126 Bari, Italy}%

\author{Oscar Straniero}\email{oscar.straniero@inaf.it}
\affiliation{INAF, Osservatorio Astronomico d’Abruzzo, 64100 Teramo, Italy}
\affiliation{Istituto Nazionale di Fisica Nucleare - Sezione di Roma, Piazzale Aldo Moro 2, 00185 Roma, Italy}%

\author{Pierluca Carenza}\email{pierluca.carenza@fysik.su.se}
\affiliation{The Oskar Klein Centre, Department of Physics, Stockholm University, AlbaNova, SE-10691 Stockholm, Sweden}

\author{Maurizio Giannotti}\email{MGiannotti@barry.edu}
\affiliation{Physical Sciences, Barry University, 11300 NE 2nd Ave., Miami Shores, FL 33161, USA}

\author{Alessandro Mirizzi}\email{alessandro.mirizzi@ba.infn.it}
\affiliation{Dipartimento Interateneo di Fisica “Michelangelo Merlin,” Via Amendola 173, 70126 Bari, Italy}
\affiliation{Istituto Nazionale di Fisica Nucleare - Sezione di Bari, Via Orabona 4, 70126 Bari, Italy}%

\date{\today}
\smallskip
\begin{abstract}
Heavy axion-like particles (ALPs), with masses up to a few 100~keV and coupled with photons can be efficiently produced in stellar plasmas, contributing to a significant energy-loss. 
This argument has been applied to helium burning stars in Globular Clusters (GCs) to obtain stringent bounds on the ALP-photon coupling $g_{a\gamma}$. 
{However, for sufficiently large values of the ALP
mass and coupling to photons, one should expect a significant fraction of ALPs to decay inside the star.}
These ALPs do not contribute to the energy loss but rather lead to an efficient energy transfer inside the star. 
We present a new \emph{ballistic} recipe that covers both the energy-loss and energy-transfer regimes and we perform the first dedicated simulation of GC stars including the  ALP energy transfer. 
This argument allows us to constrain ALPs with $m_a \lesssim 0.4$~MeV and $g_{a\gamma} \simeq 10^{-5}$~GeV$^{-1}$, probing a section of the ALP parameter space informally known as ``cosmological triangle''. This region is particularly interesting since it has been excluded only using standard cosmological arguments that can be evaded in nonstandard scenarios.
\end{abstract}
\maketitle

\emph{Introduction.---}Axions and Axion-Like Particles (ALPs) are ubiquitous in modern particle physics (see, e.g., Ref~\cite{Ringwald:2014vqa}). 
The term ALP refers usually to general pseudoscalar particles, $a$, with a two-photon vertex 
\begin{align}
    {\mathcal L}_{a\gamma}=-\frac14 g_{a\gamma} \, a\,F_{\mu\nu} \tilde{F}^{\mu\nu}\,,
\end{align}
where $a$ is the ALP field,  $F$ is the electromagnetic field strength tensor, $\tilde F$ its dual and 
$g_{a\gamma}$ is the ALP-photon coupling.
Interactions with other Standard Model fields are, in general, also possible but will not be considered in the present work.
Ultralight ALPs ($m_a \lesssim 10^{-10}$~eV~\cite{Arvanitaki:2009fg}) are especially motivated in the context of the String Axiverse~\cite{Svrcek:2006yi,Arvanitaki:2009fg,Cicoli:2012sz}.  
These, as well as other theories such as ``relaxion'' models~\cite{Graham:2015cka} 
or non-minimal QCD axion models (see, e.g., Ref.~\cite{DiLuzio:2020wdo} and references therein), predict also heavy ALPs ($m_a \gtrsim 10^{2}$~keV). 
Currently, the hunt for ALPs on this broad mass range is open through a variety of experiments with different approaches (see~\cite{Irastorza:2018dyq,Sikivie:2020zpn,Agrawal:2021dbo} for recent reviews).
In general, light ALPs can be probed by solar helioscope~\cite{CAST:2017uph,IAXO:2019mpb,BabyIAXO:2020mzw} or haloscope~\cite{ADMX:2019uok,MADMAX:2019pub} techniques, or by photon regeneration laboratory experiments~\cite{Bahre:2013ywa,OSQAR:2015qdv}. 
Instead, colliders and beam-dump experiments are capable to explore the heavy ALP mass range, reaching the
$m_a \sim {\mathcal O}$(GeV) frontier~\cite{Dolan:2017osp,Dobrich:2019dxc,Agrawal:2021dbo}.

Astrophysical arguments offer complementary opportunities to probe the ALP parameter space~\cite{Raffelt:1990yz,Raffelt:1996wa,Raffelt:2006cw,Giannotti:2015kwo,Giannotti:2017hny,DiLuzio:2021ysg}.
In particular, Globular Cluster (GC) stars have been recognised long ago as powerful astrophysical laboratories for ALPs {coupled to photons}~\cite{Raffelt:1987yu,Raffelt:1996wa,Raffelt:2006cw}. 
{Such coupling }
would allow for an efficient production in the stellar plasma, leading to an additional channel of energy-loss and thus altering the stellar evolution. 
{Consequently}, the number of stars found in the different evolutionary phases in GCs provides a valuable tool to investigate exotic energy losses in stellar interiors. In this context, the GC $R$ parameter, defined as the number ratio of horizontal branch (HB) to red giants branch (RGB) stars
\begin{equation}
R=\frac{N_{\rm HB}}{N_{\rm RGB}} \,\ , 
\end{equation}
has been used for a long time to constrain $g_{a\gamma}$. 

Light ALPs, with $m_a \lesssim 30$~keV, are produced mainly through
the Primakoff process $\gamma +Ze\to \gamma+a$, i.e. the conversion of a photon into an ALP in the electric field of ions and electrons in the stellar plasma. 
This process is considerably more efficient in HB than in RGB stars, where it is suppressed by the large plasma frequency and by electron degeneracy.
Therefore, for a sufficiently large ALP-photon coupling, the ALP emission would accelerate the stellar evolution in the HB stage, leaving the RGB phase essentially unchanged and thus leading to a reduction of the $R$ parameter. 
Comparison with the photometric data for 39 GCs lead to the bound $g_{a\gamma} \lesssim 6.6 \times 10^{-11}$~GeV$^{-1}$~\cite{Ayala:2014pea,Straniero:2015nvc}. 

The Primakoff production of heavy ALPs, with
$m_a \gtrsim 30$~keV,
is Boltzmann suppressed, so that the bound unavoidably relaxes for $m_a \gg T$.
However, the reduction of the ALP flux at large masses is partially compensated by the emergence of another axion production mechanism, the photon coalescence, $\gamma \gamma \to a$. Though, being thermal, this process suffers from the Boltzmann suppression just like the Primakoff, the steep mass dependence of the coalescence rate [see Eq.~\eqref{eq:prod}]
makes it the dominant ALP production mechanism for $m_a \gtrsim 50$~keV.
The photon coalescence process was included for the first time in the study of the HB bound on ALPs in Ref.~\cite{Carenza:2020zil}. In that study, free-streaming ALPs were included in the GC simulation as a source of energy-loss and the effect of the ALP decay, $a\to \gamma \gamma$, was accounted for only as a  reduction of the lost energy. 

A phenomenological bound was then obtained by searching for the $(m_a,\,g_{a\gamma})$ pairs for which the ALP mean free path (mfp) was smaller than the convective core.~\footnote{A similar strategy was followed in Ref.~\cite{Dolan:2021rya} to constrain ALPs using the white dwarf initial-final mass relation, obtaining bounds comparable to the one from HB stars.}
No account was given on the impact that the energy deposition within the star would have on its evolution.
However, 
for values of large enough couplings and masses a significant fraction of ALPs is expected to decay inside the star.
This effect leads to an energy transfer within the star, where ALPs are produced at a given position and deposit their energy by decay into photons at another position. 
Thus, a reliable description of the ALP impact on the evolution of HB stars cannot, in general, ignore the effects of the ALP-induced energy transport.
The energy transport in stellar interior due to exotic particles is usually described as a radiative energy transfer. For different but complementary approaches see~\cite{Raffelt:1988rx,Gould:1989ez,Gould:1989hm,Sokolov:2019cbs}.
In practice, it is treated as a diffusive phenomenon, by including an exotic component in the evaluation of the radiative opacity (Cf. Sec. 1.3.3 in Ref.~\cite{Raffelt:1996wa}). However, if the free-streaming approximation is valid when the ALP mfp is comparable or larger than the stellar radius, the diffusive approach requires an ALP mfp smaller than the characteristic temperature (or pressure) scale height.
In principle, one can treat separately free-streaming and diffusive ALP regimes, but both these assumptions fail in the case of intermediate ALP mfp. For this reason, instead of  considering two different recipes, here we propose a novel \emph{ballistic} model valid for any mfp value. An algorithm based on this model of the ALP energy transport has been included into the Full Network Stellar evolution codes (FuNS, see \cite{Straniero:2020iyi}) and used to calculate new HB stellar models.
Though we apply our strategy to the study of the impact of ALPs on the evolution of HB stars, our
method is quite general and can be adopted in other cases of exotic energy transport in stars. 
In general, one expects the ALP energy deposition to become especially  relevant for 
$m_a\sim 0.4$~MeV and $g_{a\gamma} \gtrsim 10^{-6}$~GeV$^{-1}$.
These values lay in a region, informally known as the “cosmological triangle'' ($m_{a}\sim0.5-1$~MeV and $g_{a\gamma}\sim 10^{-5}\GeV^{-1}$), which, though in tension with standard cosmological arguments~\cite{Cadamuro:2011fd,Depta:2020wmr}, is hard to access with astrophysical considerations and current experimental searches (see Ref.~\cite{Brdar:2020dpr} for a discussion about the physical potential of the planned DUNE neutrino experiment).
At small masses, the cosmological triangle is bounded by the HB bound, which we are going to revise in this paper. 
The other edges correspond to the SN 1987A bound (at small couplings) and to the experimental limits from various beam dump experiments. 
Before moving to our analysis of the HB bound, let us notice that the exact position of the SN bound, which marks the lower edge of the cosmological triangle, is also subject to uncertainties.
A recent analysis proposed that the energy deposited by decaying ALPs in the outer envelopes of the SN progenitor star must be lower than the SN explosion energy $E_{\rm SN}\sim 10^{51}$~erg.
This criterion would exclude the couplings $g_{a\gamma}\lesssim 5\times 10^{-5}$~GeV$^{-1}$ for $m_a\lesssim 10$~MeV~\cite{Caputo:2021rux},
a region large enough to cover the entire cosmological triangle. 
However, this is a semi-quantitative estimate, based on an unperturbed SN model.
Therefore, it is worthwhile to use another independent approach to probe this region. 

\emph{ALP emissivity.---} 
In this work, we are mostly concerned with massive ALPs, 
in the region of the cosmological triangle. 
As discussed above, the dominant production rate in this regime is the photon coalescence process, $\gamma\gamma\rightarrow a$ (see Ref.~\cite{Carenza:2020zil}), while the Primakoff process can be neglected.
In this case, the ALP production rate per unit volume and for ALP energy between $E$ and $E+dE$ is 

\begin{equation}
    \frac{d\dot{n}_{a}}{dE}=\frac{g_{a\gamma}^{2}}{128\pi^{3}}m_{a}^{4}p \Bigg(1-\frac{4 \omega_{\rm pl}^{2}}{m_{a}^{2}}\Bigg)^{3/2}e^{-E/T}\;,
    \label{eq:prod}
\end{equation}
where $\omega_{\rm pl}$ is the plasma frequency, $p=\sqrt{E^2-m_a^2}$ is the ALP momentum, and the photon distributions are approximated as Maxwell-Boltzmann.\footnote{The coalescence production rate for Bose-Einstein statistics has been recently provided in Ref.~\cite{Caputo:2022mah}, showing that it is larger for $m_a/T\lesssim 8$. In our case, the Maxwell-Boltzmann approximation is justified since we are considering $m_a\sim 0.4$~MeV and temperature $T\lesssim O(10)$~keV.}  In the following, the plasma frequency will be neglected since in a HB star $\omega_{\rm pl}\lesssim O(10)$~keV, much smaller than the mass $m_a>100$~keV we are interested in. 
From Eq.~\eqref{eq:prod}, 
the ALP emissivity (per unit mass) is given by the following expression
\begin{equation}
    \varepsilon_{a}=\frac{1}{\rho}\int_{m_a}^{\infty} dE\,E\,\frac{d\dot{n}_{a}}{dE}\,,
\end{equation}
where $\rho$ is the matter density.\\

\emph{ALP energy deposition.---}{ALPs produced in the stellar core may decay into photons before leaving the star, depositing energy inside it.}
This important aspect was never properly addressed in previous investigations.
Only in Ref.~\cite{Carenza:2020zil} some attempts were made to account for energy deposition in the stellar core through ALP decay, however with 
the simplified assumption that only ALPs decaying beyond the convective zone would contribute to the energy loss.
{To carry our more realistic analysis self-consistently, 
we now include the effects of the energy deposited by the decaying ALPs directly into the stellar simulations.
This allows us to check quantitatively all the outcomes of this energy deposition as well as the stellar feedback on the ALP production. }
Here we describe the ballistic method we adopt.
{We assume} that ALPs are isotropically emitted and we model the decay probability as an exponential function with a scale given by the ALP decay length 
\cite{Jaeckel:2017tud,Raffelt:2006rj}
\begin{equation}
\begin{split}
    \lambda&=\frac{64\pi}{g_{a\gamma}^2\,m_a^3}\,\frac{E}{m_a}\sqrt{1-\left(\frac{E}{m_a}\right)^{-2}}=\\
    &=0.57\, g_{5}^{-2}\,m_{100}^{-3} \frac{E}{m_a}\sqrt{1-\left(\frac{E}{m_a}\right)^{-2}}\, R_\odot,
    \end{split}
    \label{eq:mean_free_path}
\end{equation}
where $g_{5}=g_{a\gamma}/10^{-5}$~GeV$^{-1}$, $m_{100}=m_a/100$~keV 
and $R_\odot=6.957\times 10^{10}$~cm is the solar radius. 

Assuming azimuthal symmetry, for  ALPs produced at a radius $r$,  the fraction of survived particles at a radius $R$ 
after travelling   a non-radial path $l$ 
is given by $e^{-l(r,R,\alpha)/\lambda}$, 
where the path $l$ depends on the production radius $r$, the decay radius $R$ and the zenith angle $\alpha$, defined as the angle between the particle trajectory and the outward radial direction. For numerical purposes, we discretized the star envelope  in $N$ shells, each one delimited by the radii $R_{i}$ and $R_{i+1}$ ($i=1,\dots,N$, with $R_{1}=0$~km and $R_{N+1}=R_s$, being $R_s$ the star radius).
Since ALPs are  emitted isotropically, they can propagate forward ($0 \leq \alpha \leq \pi/2$) or backward ($\pi/2 < \alpha \leq \pi$). Therefore, the energy may be deposited in the $i$-th shell by ALPs produced at larger ($r > R_{i+1}$) or lower radii ($r < R_{i}$). In addition, due to the finite size of the shell, ALPs may decay in the production shell itself ($R_{i} < r < R_{i+1}$), before escaping from it. \\
The contribution $\Delta L_{i,d}$ to the rate  $\Delta L_{i}$ of energy deposited in the $i$-th shell is given by
	\begin{equation}
	\begin{split}
	\Delta L_{i,d} (\alpha)=&2\pi \int_{I_{r,d}} dr\,r^{2} \int_{m_{a}}^{\infty}dE\,E\frac{d\dot{n}_{a}(r)}{dE}\,
	\chi_d (l,\lambda)\;,
	\end{split}
	\label{eq:dL1}
	\end{equation}
where $2\pi$ comes from the integration over the azimuthal angle, $I_{r,d}$ is the integration domain for the radius, $d\dot{n}_{a}(r)/dE$ is the production rate given by Eq.~\eqref{eq:prod}, and $\chi_d (l,\lambda)$ accounts for the fraction of ALPs decaying in the $i$-th shell, depending on the path $l$ and the decay length $\lambda$. The explicit forms of $I_{r,d}$ and $\chi_d (l,\lambda)$ depend on the considered contribution.
For instance, in the case of \textit{forward emission} ($d=F$) the integration domain is $ I_{r,F} = [0, R_{i+1}]$ and
\begin{equation}
\begin{aligned}
  \chi_F &= \begin{cases}
  e^{-l(r, R_i, \alpha)/\lambda} - e^{-l(r, R_{i+1}, \alpha)/\lambda}\,, \,  &r \in [0,R_{i}) \,\ , \\
  1-e^{-l(r, R_{i+1}, \alpha)/\lambda} \, &r \in [R_{i},R_{i+1}) \,\ .

  \end{cases}
 \end{aligned}
 \end{equation}
 with the path length $l$ given by
\begin{equation}
l(r,R,\alpha)=-r\cos\alpha + R\sqrt{1-\left(\dfrac{r}{R}\right)^2\sin^2\alpha}\,.
\label{eq:l}
\end{equation}
 In Sec.~\ref{app:depos} of the Supplemental Material (SM), we provide details on the contributions related to the \textit{backward emission}.
We can compute the total rate of energy deposited in the $i$-th shell as
\begin{equation}
    \Delta L_{i} (\alpha) = \sum_{d} \Delta L_{i,d} (\alpha)\,,
\end{equation}
where the sum is over all the possible contributions.\\
The rate of energy deposited per unit mass in the $i$-th shell is defined as
\begin{equation}
\varepsilon_{{\rm dep},i} (\alpha)=\frac{\Delta L_{i}(\alpha)}{\Delta M_{i}} \,,    
\end{equation}
where $\Delta M_{i}$ is the mass enclosed in the $i$-th shell. Finally, the rate of energy deposited per unit mass averaged over the cosine of the emission angle is
given by
\begin{equation}
    \langle \varepsilon_{{\rm dep},i} \rangle = \int_0^{\pi/2} d\alpha \sin\alpha \,\ \varepsilon_{{\rm dep},i} \, ,
\label{eq:eps}
\end{equation}
where $\alpha \leq \pi/2$, with the backward emission corresponding to $\pi-\alpha$. We evaluate the integral in Eq.~\eqref{eq:eps} with a Gaussian-Legendre $N_\alpha$-point quadrature formula. 
Our results are obtained fixing $N_\alpha=10$. In Sec.~\ref{app:discretization} of the SM
we show that this choice is sufficient to guarantee a good accuracy in our numerical analysis.

\emph{ALP energy transfer in GC stars.}
The usual assumption in stellar model computations is that ALPs, once produced in the hot core, escape the star, thus acting as a local energy-loss process. 
This assumption becomes particularly inadequate if the ALP mfp is smaller than the convective core radius. In this case, the ALP production and decay processes cause an energy redistribution within the core, which reduces the temperature gradient and, in turn, limits the convective instability. 
In practice, in the case of a HB stars the ALP decay cannot be neglected for ALP masses above $m_a \sim 0.4$~MeV and coupling constants above $g_{a\gamma}\sim 10^{-6}$~GeV$^{-1}$. 
	\begin{figure}[t!]
		\vspace{0.cm}
		\includegraphics[width=0.95\columnwidth]{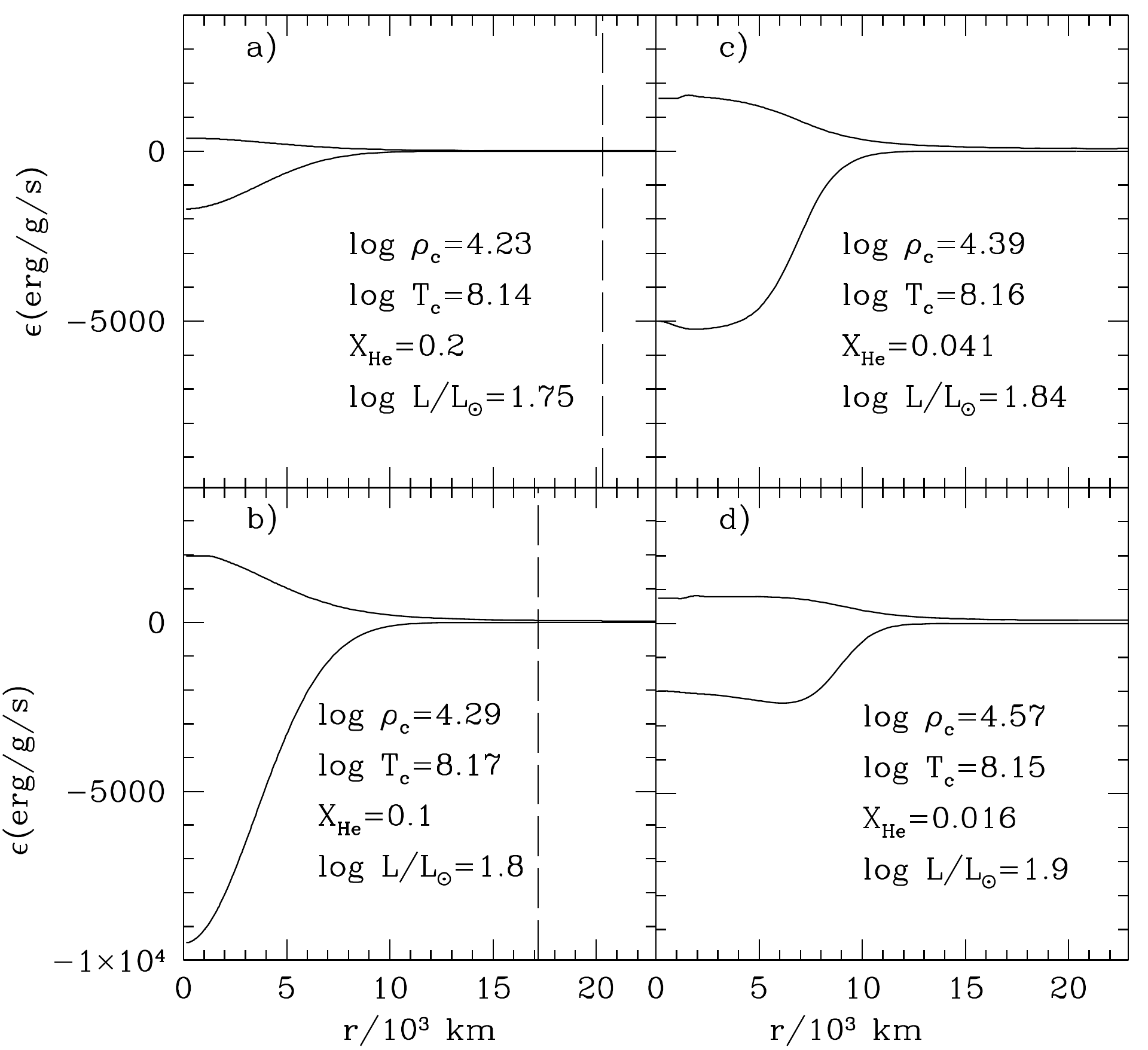}
		\caption{In each panel, the two curves show, respectively, i) the energy-loss rate, due to the coalescence process (always negative), and ii) the energy deposition rate, due to the ALP-decay process (always positive), within the core of a late HB model. The dashed vertical line in panel a) and b) marks the location of the external border of the convective core. In all panels, the central temperature, the central He mass fraction and the stellar luminosity are reported.} 
		\label{fig:energy}
	\end{figure}
In Figure \ref{fig:energy}, we show the evolution of the rate of energy loss (due to the coalescence process) and the rate of energy deposition (due to the ALP decay) within the convective core of a HB model computed assuming $m_a=0.4$~MeV and $g_{a\gamma}=3\times 10^{-6}$~GeV$^{-1}$. 
In each of the four panels, the corresponding central density and temperature,  central He mass fraction, and stellar luminosity are reported. For a large portion of the HB lifetime, the redistribution of the nuclear energy released near the centre is dominated by the convective mixing. However, when the central He mass fraction is reduced down to $X_{He}\sim0.2$, ALP production and decay start to contribute to the energy transport (panel $a$ in Fig.~\ref{fig:energy}). As a consequence, the temperature gradient becomes smaller and, in turn, the convective instability recedes. The maximum effects is attained when $X_{He}\sim0.1$ (panel $b$ in Fig.~\ref{fig:energy}). 
This causes a premature disappearance of the convective core, although the He burning is still effective near the centre.
This occurrence induces a rapid contraction of the stellar core, not coupled to an increase of the temperature, because of the combined action of ALP and plasma-neutrino production. 
As a result, the core temperature decreases slightly (the maximum $T$ moves off centre), while a substantial increase of the density occurs (panels $c$ and $d$ of Fig.~\ref{fig:energy}).

\begin{figure}[t!]
		\vspace{0.cm}
		\includegraphics[width=0.95\columnwidth]{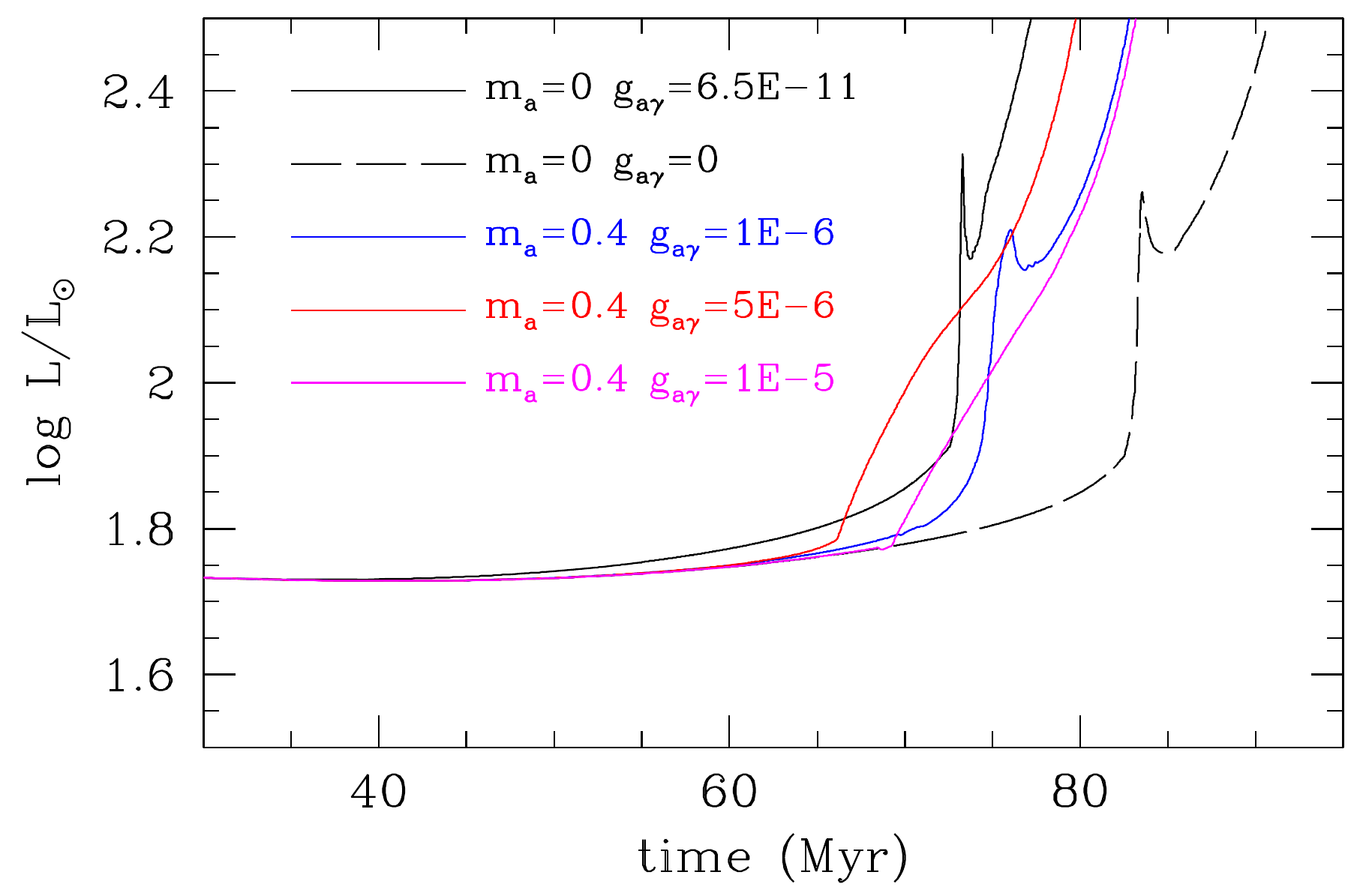}
		\caption{Luminosity versus time for HB models computed under different assumptions for ALP mass (in MeV) and coupling (in GeV$^{-1}$), as reported in the inside caption. Time 0 corresponds to the beginning of the He burning. 
		}
		\label{fig:luminosity}
	\end{figure}

In  Fig.~\ref{fig:luminosity}, we compare the luminosity evolution of HB models computed assuming different values for $m_a$ and $g_{a\gamma}$. In Ref.~\cite{Ayala:2014pea} it was shown that, assuming a conservative upper limit for the He content of the early galactic gas, $Y=0.26$, the R parameter obtained from photometric observations of 39 GCs, $R = 1.39 \pm 0.03$, implies the upper bound $g_{a\gamma} = 0.65\times 10^{-10}~\GeV^{-1}$ ($95 \%$ C.L.) for light ALPs ($m_a \lesssim 10~\keV$). 
As further discussed in Ref.~\cite{Carenza:2020zil}, in order to constrain heavier ALPs, we have evaluated the HB lifetime for a GC benchmark without exotic energy-loss (the black-dashed line in Fig.~\ref{fig:luminosity}) and for a model including light ALPs with $g_{a\gamma} = 0.65 \times 10^{-10}~\GeV^{-1}$ (black-solid line in Fig.~\ref{fig:luminosity}). 
Since the value of the R parameter is directly related to the HB lifetime, we can find the ALP bound at any mass 
in perfect analogy to what done in the case of light ALPs. Specifically, to find the $95 \%$ C.L. we should require that the HB lifetime at any fixed ALP mass is not shorter than the lifetime corresponding to a light ALP with $g_{a\gamma} = 0.65 \times 10^{-10}~\GeV^{-1}$.
The comparison with the lifetime of the reference model is done  when the stellar luminosity attains $\log L/L_\odot = 1.9$, being $L_\odot$ the Sun luminosity, a level representative of the upper HB boundary.   
According to this rule, we find that the bound in Ref.~\cite{Ayala:2014pea} for light ALPs can be reproduced for instance by assuming  $m_a=0.4$ MeV and $g_{a\gamma}=1.6\times10^{-6}$~GeV$^{-1}$. 
Smaller couplings cannot be excluded, because they lead to longer HB lifetimes. This is the case of the $m_a=0.4$ MeV and  $g_{a\gamma}=10^{-6}$~GeV$^{-1}$ model represented by the blue line in Fig.~\ref{fig:luminosity}. On the contrary, the HB is too short at larger coupling, as in the case of the model with 
 $m_a=0.4$ MeV and $g_{a\gamma}=5\times10^{-6}$~GeV$^{-1}$ represented by the red line in Fig.~\ref{fig:luminosity}. 
However, for even larger couplings (an example is the  model shown in magenta in Fig.~\ref{fig:luminosity}, with $g_{a\gamma}=1\times10^{-5}$~GeV$^{-1}$)
the HB lifetime begins to increase again.
This occurrence is due to the extreme reduction of the ALP mfp that scales as $g_{a\gamma}^{-2}$. 
Therefore, for high couplings, the  ALP mfp becomes so short that most of the ALPs decay very close to their production site 
and their contribution to the energy redistribution becomes negligible.
Thus, for each value of the ALP mass we get a pair of $g_{a\gamma}$ that reproduce the light ALP bound. 

In Sec.~\ref{app:comp} of the SM we will compare the results obtained with the ballistic method used here with the ones found using the diffusive energy-transfer approach.

\begin{figure}[t!]
		\vspace{0.cm}
		\includegraphics[width=0.95\columnwidth]{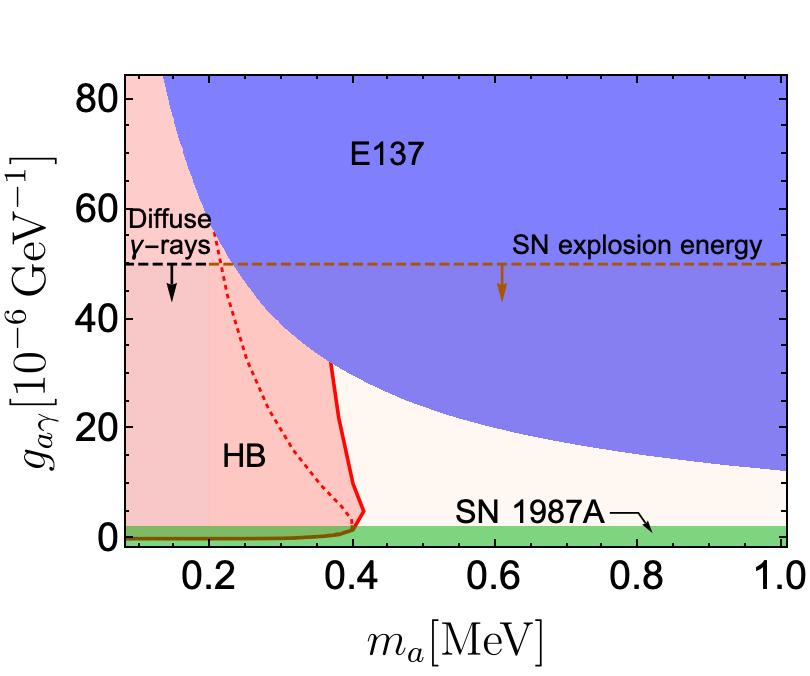}
		\caption{Overview of the parameter space around the cosmological triangle. The region shaded in light red represents the HB bound derived in this paper. The red dotted line is the previous HB bound~\cite{Carenza:2020zil}, obtained with an oversimplified recipe for the ALP decays (see text for more details). The light green region is excluded by SN 1987A~\cite{Caputo:2021rux} and the blue one by the beam dump experiment E137~\cite{Dolan:2017osp}. The SN diffuse gamma-ray background (black dashed) and the SN explosion energy (orange dashed) bounds are taken from Ref.~\cite{Caputo:2021rux}.
		}
		\label{fig:bound}
	\end{figure}

\emph{Discussion.---}
The result of our analysis is shown in Fig.~\ref{fig:bound}.
The excluded region from HB stars, derived with our novel method, is shaded in light red and delimited by the continuous red line.
The dotted line inside this region shows the previous bound, from the analysis in Ref.~\cite{Carenza:2020zil}.
Although, at a first look, it may appear that our new procedure does not change substantially the previous result, the similarity is purely accidental. In fact, the analysis in Ref.~\cite{Carenza:2020zil} is based on the crude assumption that the ALP energy loss becomes negligible when the ALP mfp is smaller than the HB convective core radius, thus neglecting effects of the ALP energy deposition and, in turn, of the consequent energy redistribution within the central convective zone.

For completeness, in the figure we also show (in light green) the region excluded by SN 1987A in the regime of ALPs trapped in the SN core~\cite{Caputo:2021rux}
and (in blue) the parameters excluded by direct searches at beam dump experiments~\cite{Dolan:2017osp,Dobrich:2019dxc,Agrawal:2021dbo}. 
As evident from the figure,
astrophysical considerations and direct searches do leave
open the region with $m_a \sim 0.5-1$~MeV and $g_{a\gamma} \simeq 10^{-5}$~GeV$^{-1}$ 
which, as discussed in the introduction, is dubbed the ``cosmological triangle''.\footnote{The cosmological triangle extends up to $m_a\sim 1.5\,$MeV (Cf. Sec.~\ref{app:comp} of the SM). 
For the sake of clarity, here we are showing only the region near the HB bound.}
Standard cosmological arguments
can constrain ALP parameters in this area. 
More specifically, this entire region is in tension with the standard Big-Bang Nucleosynthesis (BBN)  
and with considerations about the effective number of relativistic species 
$N_{\rm eff}$~\cite{Cadamuro:2011fd,Depta:2020wmr}.
However, cosmological bounds can be evaded in nonstandard cosmological histories, e.g. in low-reheating temperature models~\cite{Depta:2020wmr}.

Recently, it has been shown in Ref.~\cite{Caputo:2021rux} that  for parameters inside the cosmological triangle, ALPs produced in a SN core would fastly decay dumping all their energy into the surrounding progenitor-star matter, saturating the SN explosion energy (orange dashed band in Fig.~\ref{fig:bound}). However, a self-consistent SN simulation including ALP energy deposition, like the one we performed for HB stars, is not yet available. We hope that our approach would stimulate dedicated works also in that situation.

Before concluding, it is worth noticing a peculiar feature of models with large couplings, that could be used to get an even more stringent constraint. 
A characteristic bump is usually observed in the luminosity functions of GC AGB stars, at $\log L/L_\odot$ between 2 and 2.5. It corresponds to stars in which the H-burning shell is passing through a chemical discontinuity previously left by the receding convective envelope.
For some time, the star stops climbing the AGB and its luminosity decreases. 
Such an occurrence originates the bump observed in the GC luminosity functions.  As shown in Fig.~\ref{fig:luminosity}, this occurrence is evident at $\log L/L_\odot\sim 2.2$ in the  three models with $g_{a\gamma}=0$, $6.5\times 10^{-11}$ GeV$^{-1}$ and $10^{-6}$ GeV$^{-1}$, while it is suppressed in the two models with higher  $g_{a\gamma}$. We plan to investigate this effect in a future work.

\emph{Conclusions.---} In this Letter, we presented a detailed study of the ALP-induced energy transfer in HB stars including, for the first time, a reliable quantitative analysis of the effects of ALPs decaying into photons inside the stellar core.
For this purpose, we developed a simple recipe to model this non-local energy transfer process
and included it in our numerical simulations. 
A comparison of our recipe with the more standard
method of diffusive energy-transfer is presented in  Sec.~\ref{app:comp} of the SM.
Though applied to the specific case of HB stars in GCs, our method can be readily extended to other stars, providing a general recipe to describe energy transfer in situations in which neither the free streaming nor the diffuse approximations are fully justified.
Our study allows us the 
strengthening of the ALP-photon bound for masses $m_a\sim 0.4\,$ MeV, thus restricting significantly the ``cosmological triangle'', an area in the ALP parameter space not yet accessible to experimental searches nor to astrophysical arguments.
The cosmological triangle is a goal for several current and planned experimental searches.
A significant example is the study in Ref.~\cite{Brdar:2020dpr},
suggesting that the future DUNE experiment might be able to fully probe the ALP parameters in this area.

Our bound reduces significantly the available space for a possible discovery. 
Of course, if a positive signal were to be found in the excluded region it would have dramatic consequences on cosmology and astrophysics. Indeed, one would need to consider non-standard thermal histories to evade the cosmological bound. In conclusion the synergy between laboratory experiments, astrophysics and cosmology would be an high winning strategy to probe ALPs and might deserve unexpected surprises if discrepancies among these different approach would emerge.

\emph{Acknowlegments.---}
We warmly thank
G. Raffelt for useful comments on the manuscript.
The work of M.G. was partially supported by funding from a grant provided by the Fulbright U.S. Scholar Program.
M.G. thanks the Departamento de Física Teórica and the Centro de Astropartículas y Física de Altas Energías (CAPA) of the Universidad de Zaragoza for hospitality during the completion of a large part of this work. For this work, O.S has been funded by the Italian Space Agency (ASI) and the Italian National Institute of Astrophysics (INAF) under the agreement n. 2017-14-H.0 -attività di studio per la comunità scientifica di Astrofisica delle AlteEnergie e Fisica
Astroparticellare.
The work of G.L. and 
A.M. is partially supported by the Italian Istituto Nazionale di Fisica Nucleare (INFN) through the ``Theoretical Astroparticle Physics'' project
and by the research grant number 2017W4HA7S
``NAT-NET: Neutrino and Astroparticle Theory Network'' under the program
PRIN 2017 funded by the Italian Ministero dell'Universit\`a e della
Ricerca (MUR). 
The work of P.C. is supported by the European Research Council under Grant No. 742104 and by
the Swedish Research Council (VR) under grants 2018-03641 and 2019-02337. 

\bibliographystyle{bibi.bst}
\bibliography{biblio.bib}

\providecommand{\href}[2]{#2}\begingroup\raggedright\begin{thebibliography}{10}

\bibitem{Ringwald:2014vqa}
A.~Ringwald, \emph{{Axions and Axion-Like Particles}},  in \emph{{49th
  Rencontres de Moriond on Electroweak Interactions and Unified Theories}}, 7,
  2014, \href{https://arxiv.org/abs/1407.0546}{{\ttfamily 1407.0546}}.

\bibitem{Arvanitaki:2009fg}
A.~Arvanitaki, S.~Dimopoulos, S.~Dubovsky, N.~Kaloper and J.~March-Russell,
  \emph{{String Axiverse}},
  \href{https://doi.org/10.1103/PhysRevD.81.123530}{\emph{Phys. Rev. D}
  {\bfseries 81} (2010) 123530}
  [\href{https://arxiv.org/abs/0905.4720}{{\ttfamily 0905.4720}}].

\bibitem{Svrcek:2006yi}
P.~Svrcek and E.~Witten, \emph{{Axions In String Theory}},
  \href{https://doi.org/10.1088/1126-6708/2006/06/051}{\emph{JHEP} {\bfseries
  06} (2006) 051} [\href{https://arxiv.org/abs/hep-th/0605206}{{\ttfamily
  hep-th/0605206}}].

\bibitem{Cicoli:2012sz}
M.~Cicoli, M.~Goodsell and A.~Ringwald, \emph{{The type IIB string axiverse and
  its low-energy phenomenology}},
  \href{https://doi.org/10.1007/JHEP10(2012)146}{\emph{JHEP} {\bfseries 10}
  (2012) 146} [\href{https://arxiv.org/abs/1206.0819}{{\ttfamily 1206.0819}}].

\bibitem{Graham:2015cka}
P.~W. Graham, D.~E. Kaplan and S.~Rajendran, \emph{{Cosmological Relaxation of
  the Electroweak Scale}},
  \href{https://doi.org/10.1103/PhysRevLett.115.221801}{\emph{Phys. Rev. Lett.}
  {\bfseries 115} (2015) 221801}
  [\href{https://arxiv.org/abs/1504.07551}{{\ttfamily 1504.07551}}].

\bibitem{DiLuzio:2020wdo}
L.~Di~Luzio, M.~Giannotti, E.~Nardi and L.~Visinelli, \emph{{The landscape of
  QCD axion models}},
  \href{https://doi.org/10.1016/j.physrep.2020.06.002}{\emph{Phys. Rept.}
  {\bfseries 870} (2020) 1} [\href{https://arxiv.org/abs/2003.01100}{{\ttfamily
  2003.01100}}].

\bibitem{Irastorza:2018dyq}
I.~G. Irastorza and J.~Redondo, \emph{{New experimental approaches in the
  search for axion-like particles}},
  \href{https://doi.org/10.1016/j.ppnp.2018.05.003}{\emph{Prog. Part. Nucl.
  Phys.} {\bfseries 102} (2018) 89}
  [\href{https://arxiv.org/abs/1801.08127}{{\ttfamily 1801.08127}}].

\bibitem{Sikivie:2020zpn}
P.~Sikivie, \emph{{Invisible Axion Search Methods}},
  \href{https://doi.org/10.1103/RevModPhys.93.015004}{\emph{Rev. Mod. Phys.}
  {\bfseries 93} (2021) 015004}
  [\href{https://arxiv.org/abs/2003.02206}{{\ttfamily 2003.02206}}].

\bibitem{Agrawal:2021dbo}
P.~Agrawal et~al., \emph{{Feebly-interacting particles: FIPs 2020 workshop
  report}}, \href{https://doi.org/10.1140/epjc/s10052-021-09703-7}{\emph{Eur.
  Phys. J. C} {\bfseries 81} (2021) 1015}
  [\href{https://arxiv.org/abs/2102.12143}{{\ttfamily 2102.12143}}].

\bibitem{CAST:2017uph}
{\scshape CAST} Collaboration, V.~Anastassopoulos et~al., \emph{{New CAST Limit
  on the Axion-Photon Interaction}},
  \href{https://doi.org/10.1038/nphys4109}{\emph{Nature Phys.} {\bfseries 13}
  (2017) 584} [\href{https://arxiv.org/abs/1705.02290}{{\ttfamily
  1705.02290}}].

\bibitem{IAXO:2019mpb}
{\scshape IAXO} Collaboration, E.~Armengaud et~al., \emph{{Physics potential of
  the International Axion Observatory (IAXO)}},
  \href{https://doi.org/10.1088/1475-7516/2019/06/047}{\emph{JCAP} {\bfseries
  06} (2019) 047} [\href{https://arxiv.org/abs/1904.09155}{{\ttfamily
  1904.09155}}].

\bibitem{BabyIAXO:2020mzw}
{\scshape BabyIAXO} Collaboration, A.~Abeln et~al., \emph{{Conceptual Design of
  BabyIAXO, the intermediate stage towards the International Axion
  Observatory}},  \href{https://arxiv.org/abs/2010.12076}{{\ttfamily
  2010.12076}}.

\bibitem{ADMX:2019uok}
{\scshape ADMX} Collaboration, T.~Braine et~al., \emph{{Extended Search for the
  Invisible Axion with the Axion Dark Matter Experiment}},
  \href{https://doi.org/10.1103/PhysRevLett.124.101303}{\emph{Phys. Rev. Lett.}
  {\bfseries 124} (2020) 101303}
  [\href{https://arxiv.org/abs/1910.08638}{{\ttfamily 1910.08638}}].

\bibitem{MADMAX:2019pub}
{\scshape MADMAX} Collaboration, P.~Brun et~al., \emph{{A new experimental
  approach to probe QCD axion dark matter in the mass range above 40 $\mu$eV}},
  \href{https://doi.org/10.1140/epjc/s10052-019-6683-x}{\emph{Eur. Phys. J. C}
  {\bfseries 79} (2019) 186}
  [\href{https://arxiv.org/abs/1901.07401}{{\ttfamily 1901.07401}}].

\bibitem{Bahre:2013ywa}
R.~B\"ahre et~al., \emph{{Any light particle search II \textemdash{}Technical
  Design Report}},
  \href{https://doi.org/10.1088/1748-0221/8/09/T09001}{\emph{JINST} {\bfseries
  8} (2013) T09001} [\href{https://arxiv.org/abs/1302.5647}{{\ttfamily
  1302.5647}}].

\bibitem{OSQAR:2015qdv}
{\scshape OSQAR} Collaboration, R.~Ballou et~al., \emph{{New exclusion limits
  on scalar and pseudoscalar axionlike particles from light shining through a
  wall}}, \href{https://doi.org/10.1103/PhysRevD.92.092002}{\emph{Phys. Rev. D}
  {\bfseries 92} (2015) 092002}
  [\href{https://arxiv.org/abs/1506.08082}{{\ttfamily 1506.08082}}].

\bibitem{Dolan:2017osp}
M.~J. Dolan, T.~Ferber, C.~Hearty, F.~Kahlhoefer and K.~Schmidt-Hoberg,
  \emph{{Revised constraints and Belle II sensitivity for visible and invisible
  axion-like particles}},
  \href{https://doi.org/10.1007/JHEP12(2017)094}{\emph{JHEP} {\bfseries 12}
  (2017) 094} [\href{https://arxiv.org/abs/1709.00009}{{\ttfamily
  1709.00009}}]. [Erratum: JHEP 03, 190 (2021)].

\bibitem{Dobrich:2019dxc}
B.~D\"obrich, J.~Jaeckel and T.~Spadaro, \emph{{Light in the beam dump - ALP
  production from decay photons in proton beam-dumps}},
  \href{https://doi.org/10.1007/JHEP05(2019)213}{\emph{JHEP} {\bfseries 05}
  (2019) 213} [\href{https://arxiv.org/abs/1904.02091}{{\ttfamily
  1904.02091}}]. [Erratum: JHEP 10, 046 (2020)].

\bibitem{Raffelt:1990yz}
G.~G. Raffelt, \emph{{Astrophysical methods to constrain axions and other novel
  particle phenomena}},
  \href{https://doi.org/10.1016/0370-1573(90)90054-6}{\emph{Phys. Rept.}
  {\bfseries 198} (1990) 1}.

\bibitem{Raffelt:1996wa}
G.~G. Raffelt, \emph{{Stars as laboratories for fundamental physics}}. Chicago,
  USA: Univ. Pr., 1996.

\bibitem{Raffelt:2006cw}
G.~G. Raffelt, \emph{{Astrophysical axion bounds}},
  \href{https://doi.org/10.1007/978-3-540-73518-2_3}{\emph{Lect. Notes Phys.}
  {\bfseries 741} (2008) 51}
  [\href{https://arxiv.org/abs/hep-ph/0611350}{{\ttfamily hep-ph/0611350}}].

\bibitem{Giannotti:2015kwo}
M.~Giannotti, I.~Irastorza, J.~Redondo and A.~Ringwald, \emph{{Cool WISPs for
  stellar cooling excesses}},
  \href{https://doi.org/10.1088/1475-7516/2016/05/057}{\emph{JCAP} {\bfseries
  05} (2016) 057} [\href{https://arxiv.org/abs/1512.08108}{{\ttfamily
  1512.08108}}].

\bibitem{Giannotti:2017hny}
M.~Giannotti, I.~G. Irastorza, J.~Redondo, A.~Ringwald and K.~Saikawa,
  \emph{{Stellar Recipes for Axion Hunters}},
  \href{https://doi.org/10.1088/1475-7516/2017/10/010}{\emph{JCAP} {\bfseries
  10} (2017) 010} [\href{https://arxiv.org/abs/1708.02111}{{\ttfamily
  1708.02111}}].

\bibitem{DiLuzio:2021ysg}
L.~Di~Luzio, M.~Fedele, M.~Giannotti, F.~Mescia and E.~Nardi, \emph{{Stellar
  evolution confronts axion models}},
  \href{https://doi.org/10.1088/1475-7516/2022/02/035}{\emph{JCAP} {\bfseries
  02} (2022) 035} [\href{https://arxiv.org/abs/2109.10368}{{\ttfamily
  2109.10368}}].

\bibitem{Raffelt:1987yu}
G.~G. Raffelt and D.~S.~P. Dearborn, \emph{{Bounds on Hadronic Axions From
  Stellar Evolution}},
  \href{https://doi.org/10.1103/PhysRevD.36.2211}{\emph{Phys. Rev. D}
  {\bfseries 36} (1987) 2211}.

\bibitem{Ayala:2014pea}
A.~Ayala, I.~Dom\'\i{}nguez, M.~Giannotti, A.~Mirizzi and O.~Straniero,
  \emph{{Revisiting the bound on axion-photon coupling from Globular
  Clusters}}, \href{https://doi.org/10.1103/PhysRevLett.113.191302}{\emph{Phys.
  Rev. Lett.} {\bfseries 113} (2014) 191302}
  [\href{https://arxiv.org/abs/1406.6053}{{\ttfamily 1406.6053}}].

\bibitem{Straniero:2015nvc}
O.~Straniero, A.~Ayala, M.~Giannotti, A.~Mirizzi and I.~Dominguez,
  \emph{{Axion-Photon Coupling: Astrophysical Constraints}},  in \emph{{11th
  Patras Workshop on Axions, WIMPs and WISPs}}, pp.~77--81, 2015,
  \href{https://doi.org/10.3204/DESY-PROC-2015-02/straniero_oscar}{DOI}.

\bibitem{Carenza:2020zil}
P.~Carenza, O.~Straniero, B.~D\"obrich, M.~Giannotti, G.~Lucente and
  A.~Mirizzi, \emph{{Constraints on the coupling with photons of heavy
  axion-like-particles from Globular Clusters}},
  \href{https://doi.org/10.1016/j.physletb.2020.135709}{\emph{Phys. Lett. B}
  {\bfseries 809} (2020) 135709}
  [\href{https://arxiv.org/abs/2004.08399}{{\ttfamily 2004.08399}}].

\bibitem{Dolan:2021rya}
M.~J. Dolan, F.~J. Hiskens and R.~R. Volkas, \emph{{Constraining axion-like
  particles using the white dwarf initial-final mass relation}},
  \href{https://doi.org/10.1088/1475-7516/2021/09/010}{\emph{JCAP} {\bfseries
  09} (2021) 010} [\href{https://arxiv.org/abs/2102.00379}{{\ttfamily
  2102.00379}}].

\bibitem{Raffelt:1988rx}
G.~G. Raffelt and G.~D. Starkman, \emph{{STELLAR ENERGY TRANSFER BY keV MASS
  SCALARS}}, \href{https://doi.org/10.1103/PhysRevD.40.942}{\emph{Phys. Rev. D}
  {\bfseries 40} (1989) 942}.

\bibitem{Gould:1989ez}
A.~Gould and G.~Raffelt, \emph{{Cosmion Energy Transfer in Stars: The Knudsen
  Limit}}, \href{https://doi.org/10.1086/168569}{\emph{Astrophys. J.}
  {\bfseries 352} (1990) 669}.

\bibitem{Gould:1989hm}
A.~Gould and G.~Raffelt, \emph{{THERMAL CONDUCTION BY MASSIVE PARTICLES}},
  \href{https://doi.org/10.1086/168568}{\emph{Astrophys. J.} {\bfseries 352}
  (1990) 654}.

\bibitem{Sokolov:2019cbs}
A.~V. Sokolov, \emph{{Generic energy transport solutions to the solar abundance
  problem\textemdash{}a hint of new physics}},
  \href{https://doi.org/10.1088/1475-7516/2020/03/013}{\emph{JCAP} {\bfseries
  03} (2020) 013} [\href{https://arxiv.org/abs/1907.06928}{{\ttfamily
  1907.06928}}].

\bibitem{Straniero:2020iyi}
O.~Straniero, C.~Pallanca, E.~Dalessandro, I.~Dominguez, F.~R. Ferraro,
  M.~Giannotti, A.~Mirizzi and L.~Piersanti, \emph{{The RGB tip of galactic
  globular clusters and the revision of the bound of the axion-electron
  coupling}}, \href{https://doi.org/10.1051/0004-6361/202038775}{\emph{Astron.
  Astrophys.} {\bfseries 644} (2020) A166}
  [\href{https://arxiv.org/abs/2010.03833}{{\ttfamily 2010.03833}}].

\bibitem{Cadamuro:2011fd}
D.~Cadamuro and J.~Redondo, \emph{{Cosmological bounds on pseudo
  Nambu-Goldstone bosons}},
  \href{https://doi.org/10.1088/1475-7516/2012/02/032}{\emph{JCAP} {\bfseries
  02} (2012) 032} [\href{https://arxiv.org/abs/1110.2895}{{\ttfamily
  1110.2895}}].

\bibitem{Depta:2020wmr}
P.~F. Depta, M.~Hufnagel and K.~Schmidt-Hoberg, \emph{{Robust cosmological
  constraints on axion-like particles}},
  \href{https://doi.org/10.1088/1475-7516/2020/05/009}{\emph{JCAP} {\bfseries
  05} (2020) 009} [\href{https://arxiv.org/abs/2002.08370}{{\ttfamily
  2002.08370}}].

\bibitem{Brdar:2020dpr}
V.~Brdar, B.~Dutta, W.~Jang, D.~Kim, I.~M. Shoemaker, Z.~Tabrizi, A.~Thompson
  and J.~Yu, \emph{{Axionlike Particles at Future Neutrino Experiments: Closing
  the Cosmological Triangle}},
  \href{https://doi.org/10.1103/PhysRevLett.126.201801}{\emph{Phys. Rev. Lett.}
  {\bfseries 126} (2021) 201801}
  [\href{https://arxiv.org/abs/2011.07054}{{\ttfamily 2011.07054}}].

\bibitem{Caputo:2021rux}
A.~Caputo, G.~Raffelt and E.~Vitagliano, \emph{{Muonic boson limits: Supernova
  redux}}, \href{https://doi.org/10.1103/PhysRevD.105.035022}{\emph{Phys. Rev.
  D} {\bfseries 105} (2022) 035022}
  [\href{https://arxiv.org/abs/2109.03244}{{\ttfamily 2109.03244}}].

\bibitem{Caputo:2022mah}
A.~Caputo, H.-T. Janka, G.~Raffelt and E.~Vitagliano, \emph{{Low-Energy
  Supernovae Severely Constrain Radiative Particle Decays}},
  \href{https://doi.org/10.1103/PhysRevLett.128.221103}{\emph{Phys. Rev. Lett.}
  {\bfseries 128} (2022) 221103}
  [\href{https://arxiv.org/abs/2201.09890}{{\ttfamily 2201.09890}}].

\bibitem{Jaeckel:2017tud}
J.~Jaeckel, P.~C. Malta and J.~Redondo, \emph{{Decay photons from the axionlike
  particles burst of type II supernovae}},
  \href{https://doi.org/10.1103/PhysRevD.98.055032}{\emph{Phys. Rev. D}
  {\bfseries 98} (2018) 055032}
  [\href{https://arxiv.org/abs/1702.02964}{{\ttfamily 1702.02964}}].

\bibitem{Raffelt:2006rj}
G.~G. Raffelt, \emph{{Axions: Motivation, limits and searches}},
  \href{https://doi.org/10.1088/1751-8113/40/25/S05}{\emph{J. Phys. A}
  {\bfseries 40} (2007) 6607}
  [\href{https://arxiv.org/abs/hep-ph/0611118}{{\ttfamily hep-ph/0611118}}].

\bibitem{KW1990}
R.~{Kippenhahn} and A.~{Weigert}, \emph{{Stellar Structure and Evolution}}.
  Springer-Verlag Berlin Heidelberg New York.~Also Astronomy and Astrophysics
  Library, 1990.

\end{thebibliography}\endgroup

\clearpage

\onecolumngrid
\begin{center}
  \textbf{\large Supplemental Material: Constraining heavy axion-like particles by  energy deposition  in Globular Cluster stars}
\end{center}

\twocolumngrid
\setcounter{equation}{0}
\setcounter{figure}{0}
\setcounter{table}{0}
\setcounter{section}{0}
\setcounter{page}{1}
\makeatletter
\renewcommand{\theequation}{S\arabic{equation}}
\renewcommand{\thefigure}{S\arabic{figure}}
\renewcommand{\thetable}{S\arabic{table}}

\onecolumngrid

\section{Description of the ballistic method}
\label{app:method}

\subsection{Energy deposition}
\label{app:depos}
Since ALPs are isotropically emitted, in the shell $[R_i;\,R_{i+1}]$ the energy can be deposited by the decay of ALPs produced in outer ($r > R_{i+1}$) or inner ($r < R_{i}$) shells. In addition, due to the finite size of the shell, ALPs may decay in the production shell itself ($R_{i} < r < R_{i+1}$), before escaping it. Assuming azimuthal symmetry, the path $l$ followed by the decaying ALPs depends on the production radius $r$, on the decay radius $R$ and the zenith angle $\alpha$, defined as the angle between the particle trajectory and the outward radial direction. 
ALPs propagate forward for $0 \leq \alpha \leq \pi/2$, and backward for $\pi/2 < \alpha \leq \pi$. In order to evaluate the impact of the energy transferred by decaying ALPs we need to evaluate the rate of energy deposited in the $i$-th shell, averaged over the cosine of the zenith angle $\alpha$. We take the 
angle $\alpha \in [0;\pi/2]$ so that the backward emission corresponds  to $\pi-\alpha$.
Therefore the rate of energy deposited per unit mass, averaged over the cosine of the zenith angle $\alpha$, is given by

\begin{equation}
    \langle \varepsilon_{{\rm dep}} \rangle = \frac{\int_0^{\pi/2} d\alpha \sin\alpha \,\ \varepsilon_{{\rm dep}} (\alpha)}{\int_0^{\pi/2} d\alpha \sin\alpha} = \int_{0}^{\pi/2} d\alpha \sin\alpha \varepsilon_{\rm dep} (\alpha)\,.
    \label{eq:aveeps}
\end{equation}

\begin{figure*}[t!]
	\centering
	\includegraphics[scale=0.4]{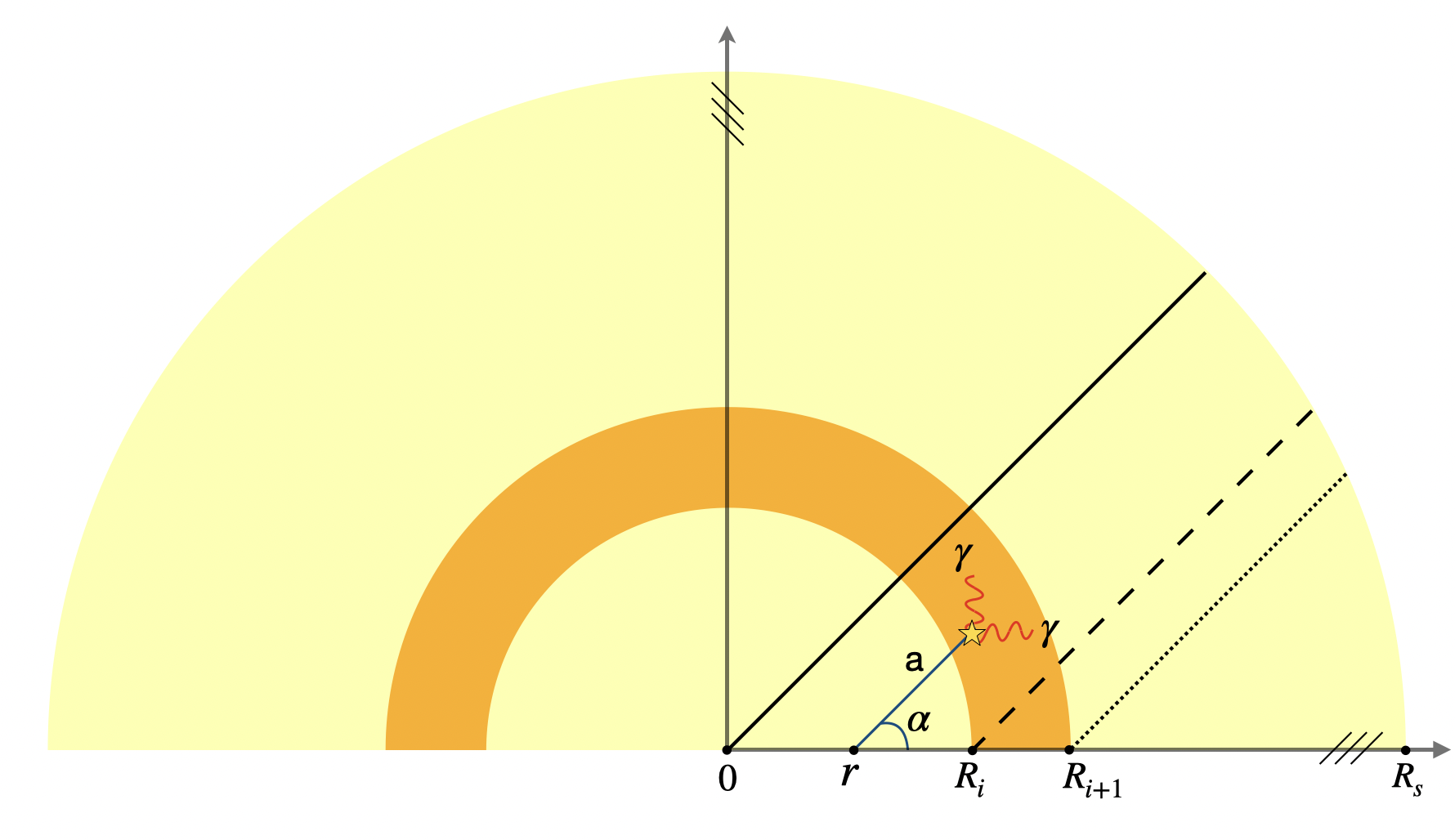}   				
	\caption{Geometrical representation of ALP forward emission, at fixed emission angle $\alpha$. The $i$-th shell is the orange area. The black lines represent the trajectories delimiting the production zones in Eq.~\eqref{eq:FS}.  The picture is not to scale.}
	\label{fig:forw}
\end{figure*}

We evaluate the integral in Eq.~\eqref{eq:aveeps} with a Gaussian-Legendre $N_\alpha$-point quadrature formula. At fixed emission angle $\alpha$, $\varepsilon_{\rm dep} (\alpha)$, the rate of energy deposited in the $i$-th shell supposing that half of the ALPs are forward emitted at $\alpha$ and the other half backward at $\pi-\alpha$, is given by
\begin{equation}
\varepsilon_{{\rm dep},i} (\alpha)=\frac{\Delta L_{i}(\alpha)}{\Delta M_{i}} \,,    
\end{equation}
where  $\Delta M_{i}$ is the mass enclosed in the $i$-th shell and $\Delta L_{i}(\alpha)$ is the rate of the energy deposited in the $i$-th shell, evaluated as the sum of different contributions $	\Delta L_{i,d} (\alpha)$. Each contribution is given by
	\begin{equation}
	\begin{split}
	\Delta L_{i,d} (\alpha)=&2\pi \int_{I_{r,d}} dr\,r^{2} \int_{m_{a}}^{\infty}dE\,E\frac{d\dot{n}_{a}(r)}{dE}\,
	\chi_d (l,\lambda)\;,
	\end{split}
	\label{eq:DLi}
	\end{equation}
where $2\pi$ comes from the integration over the azimuthal angle, $d\dot{n}_{a}(r)/dE$ is the production rate given by Eq.~\eqref{eq:prod}, $I_{r,d}$ is the integration domain for the production radius and $\chi_d (l,\lambda)$ accounts for the fraction of ALPs decaying in the $i$-th shell, depending on the path $l$ and the decay length $\lambda$. The explicit expressions of $I_{r,d}$, $\chi_d (l,\lambda)$ and $l$ depend on the considered contribution, with $l$ assuming only two possible forms
\begin{equation}
l_{\pm}(r,R,\beta)=-r\cos\beta \pm R\sqrt{1-\left(\dfrac{r}{R}\right)^2\sin^2\beta}\,,
\label{eq:lpm}
\end{equation}
where $\beta = \alpha $ in the case of forward emission and $\beta = \pi-\alpha $ for backward emission. We stress that $l_+>0$ if $\cos\beta<0$ (for any value of $r$ and $R$) or $\cos\beta>0$ and $r<R$, while $l_->0$ only for $\cos\beta<0$ and $R<r<R/\sin\beta$.
In order to characterise the possible contributions to the energy deposition, two auxiliary functions are introduced
\begin{equation}
    f_\pm (\beta) = e^{-l_\pm (r,R_i,\beta)/\lambda} - e^{-l_\pm (r,R_{i+1},\beta)/\lambda} \quad  \textrm{and} \quad    g_{\pm,i} (\beta) = 1 - e^{-l_\pm (r,R_i,\beta)/\lambda}\,,
\end{equation}
where $R_{i}$ and $R_{i+1}$ are respectively the lower and the upper boundary radius  of the $i$-th shell and the subscript $i$ for the function $g$ refers to $R_{i}$.
In the following, we will give details about each contribution, distinguishing between ALPs \emph{forward} and \emph{backward} emitted.    \\
In the case of forward emission (see Fig.~\ref{fig:forw}), the contribution to the energy deposited in the $i$-th shell is given by Eq.~\eqref{eq:DLi}, with $I_{r,F}=[0,R_{i+1}]$ and

\begin{equation}
\begin{aligned}
  \chi_F &= \begin{cases}
  f_+(\alpha) \qquad  &r \in [0,R_{i}) \,\ , \\
  g_{+,i+1}(\alpha) \qquad  &r \in [R_{i},R_{i+1}) \,\ .

  \end{cases}
 \end{aligned}
 \label{eq:FS}
 \end{equation}
 
\begin{figure*}[t!!]
	\centering
	\includegraphics[scale=0.4]{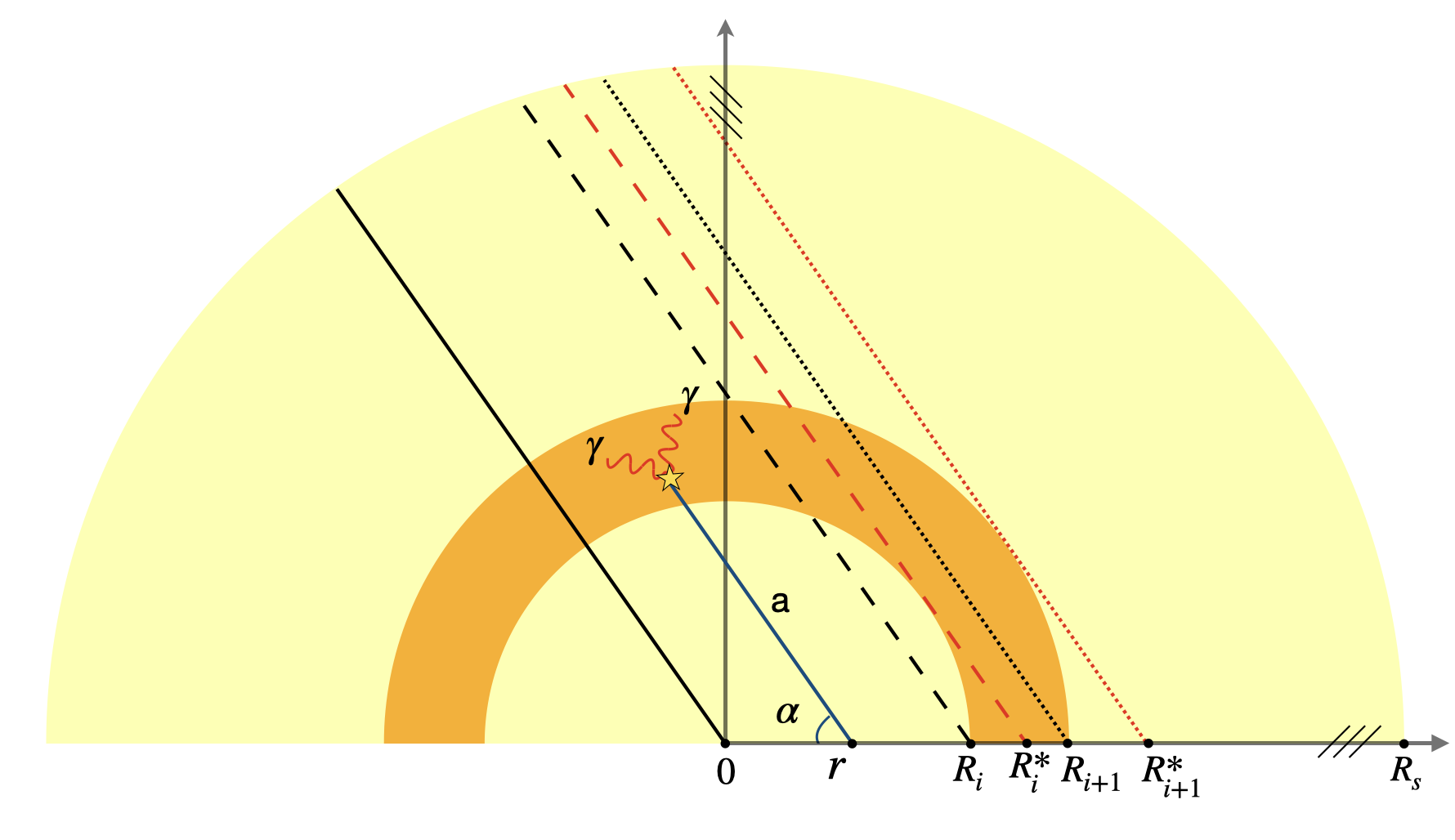}  
	\includegraphics[scale=0.4]{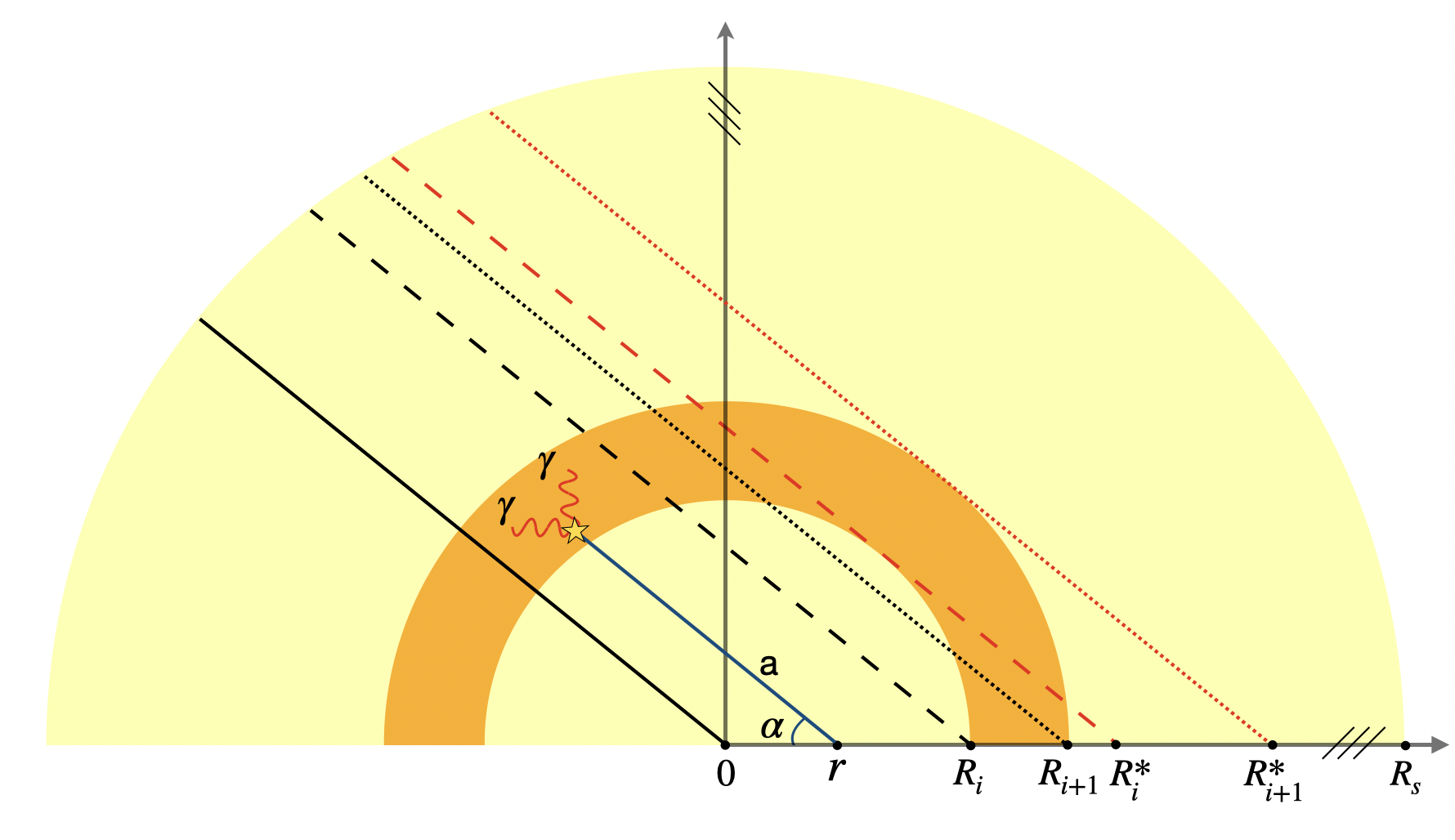}   
	\caption{Geometrical representation of ALPs backward-emitted at fixed $\alpha$. The $i$-th shell is the orange area, while the straight lines represent the trajectories delimiting the production zones in Eqs.~\eqref{eq:backw<} and \eqref{eq:backw>}. We show in red the trajectories starting from $R_i^*$ and $R_{i+1}^*$, whose position distinguishes the two cases: $R_{i}^*<R_{i+1}$ (upper panel) and $R_{i}^*>R_{i+1}$ (lower panel). The picture is not to scale.}
	\label{fig:backw}
\end{figure*}

The backward emission case (see Fig.~\ref{fig:backw}) is trickier. When ALPs are produced backward from inner shells ($r<R_{i}$), they always have a chance to cross the $i$-th shell (unless they decay before reaching it). On the other hand, more attention is needed to consider the energy deposited by ALPs backward emitted from outer shells. Indeed, an ALP backward emitted with zenith angle $\pi-\alpha$ at a radius $r$ intersects the sphere of radius $R_i<r$ at most twice, after covering a path of length $l_-(r,R_i,\pi-\alpha)$ at the first crossing and $l_+(r,R_i,\pi-\alpha)$ at the second intersection. An analogous situation occurs for the crossing of the sphere with radius $R_{i+1}$. From Eq.~\eqref{eq:lpm} we note that, for a given backward emission angle $\pi-\alpha$ there is a maximum radius $R_i^*=R_i/\sin\alpha$ for which there is an intersection tangent to the sphere of radius $R_i$ and analogously $R_{i+1}^*=R_{i+1}/\sin\alpha$ for a crossing tangent to the sphere of radius $R_{i+1}$. The contributions to the rate of deposited energy depend on the value of $R_{i}^*$ with respect to $R_{i+1}$.\\
In particular, if $R_{i}^*<R_{i+1}$ (see the upper panel in Fig.~\ref{fig:backw}) the contribution to the energy deposited in the $i$-th shell $\Delta L_{i,B_1}$ is characterized by $I_{r,B_{1}}=[0,R_{i+1}^*]$ and

\begin{equation}\label{eq:backw<}
\begin{aligned}
  \chi_{B_1} &= \begin{cases}
 f_+(\pi-\alpha) \qquad&  r \in [0,R_{i}] \,\ , \\
 g_{-,i}(\pi-\alpha)+f_{+}(\pi-\alpha) \qquad&  r \in (R_{i},R_{i}^*] \,\ , \\
 g_{+,i+1}(\pi-\alpha) \qquad&  r \in (R_{i}^*,R_{i+1}] \,\ , \\
 g_{+,i+1}(\pi-\alpha)-g_{-,i+1}(\pi-\alpha) \qquad&  r \in (R_{i+1},R_{i+1}^*] \,\ . 
  \end{cases}
 \end{aligned}
 \end{equation}
 \\
On the other hand, if $R_{i}^*>R_{i+1}$ (see the lower panel in Fig.~\ref{fig:backw}), the contribution to the energy deposited in the $i$-th shell $\Delta\,L_{i,B_2}$ is characterized by $I_{r,B_{2}}=[0,R_{i+1}^*]$ and

\begin{equation}\label{eq:backw>}
\begin{aligned}
  \chi_{B_2} &= \begin{cases}
 f_+(\pi-\alpha) \qquad &  r \in [0,R_{i}] \,\ , \\ 
 g_{-,i}(\pi-\alpha)+f_{+}(\pi-\alpha) \qquad &  r \in (R_{i},R_{i+1}] \,\ , \\ 
 f_{+}(\pi-\alpha)-f_{-}(\pi-\alpha) \qquad &  r \in (R_{i+1},R_{i}^*] \,\ ,\\ 
 g_{+,i+1}(\pi-\alpha)-g_{-,i+1}(\pi-\alpha) \qquad &  r \in (R_{i}^*,R_{i+1}^*] \,\ .
  \end{cases}
 \end{aligned}
 \end{equation}
 
Given the previously described contributions, the rate of energy deposited in the $i$-th shell is given by
\begin{equation}
    \Delta L_{i} = \sum_{d} \Delta L_{i,d}\,,
\end{equation}
where the sum is over all the possible contributions, i.e. $d=F,\,B_1$ if $R_{i}^*<R_{i+1}$ and $d=F,\,B_2$ if $R_{i}^*>R_{i+1}$.

\subsection{Effects of the discretization}
\label{app:discretization}
\begin{figure*}[t!]
	\centering
	\includegraphics[scale=0.38]{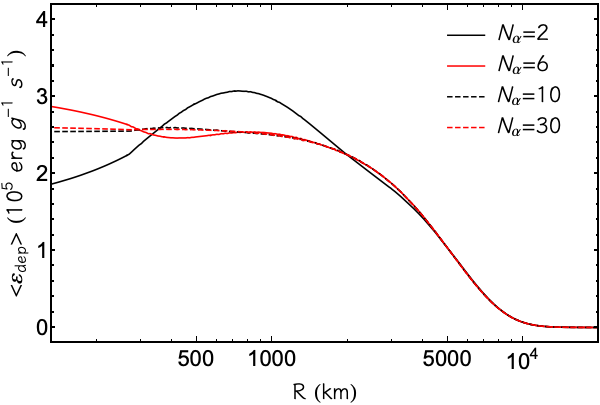}   	
	\includegraphics[scale=0.39]{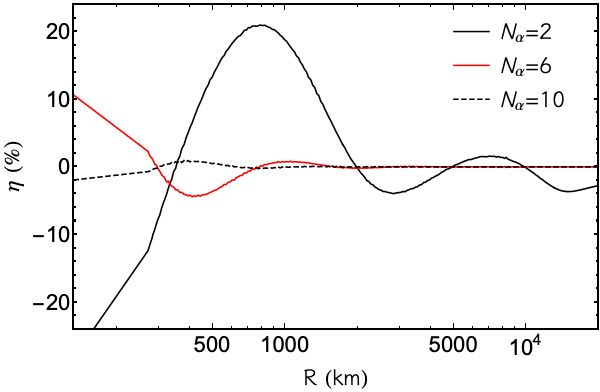}  
	\caption{The rate of energy deposited per unit mass (\emph{upper panel}) and the percentage relative error in Eq.~\eqref{eq:eta} (\emph{lower panel}) as a function of the stellar radius $R$, for different values of $N_\alpha$ as shown in the legend. }
	\label{fig:discr}
\end{figure*}

In this Section, the impact of the discretization in the Gaussian-Legendre quadrature formula is discussed. Since the computation time scales linearly with the considered emission angles $N_\alpha$, a compromise between computational time and precision of the code is needed to be found. 
We numerically checked that on an unperturbed model, independently of ALP coupling and mass, the net energy flux (i.e. the difference between the emitted energy and the sum of the deposited and lost energy) is zero within $0.1\%$ for each value of $N_\alpha$. Although the energy is conserved, the discretization affects the spacial distribution of the energy deposited throughout the star. Indeed, as shown in the left panel of Fig.~\ref{fig:discr} for reference values $m_a=0.4$~MeV and $g_{a\gamma}=10^{-5}$ GeV$^{-1}$ (corresponding to  decay length $\lambda \approx O(10^3)$~km), for low values of $N_\alpha$ ($N_\alpha \lesssim 10$) there is a fictitious off-centre peak in $\langle \varepsilon_{\rm dep} \rangle$, which disappears in the continuum limit $N_\alpha \rightarrow \infty$ (we show $\langle \varepsilon_{\rm dep} \rangle$ for $N_\alpha=30$ as an example of this limit).
This fictitious off-centre peak is related to the fact that, as discussed in Sec.~\ref{app:depos}, ALPs produced with emission angle $\alpha$ at a radius $r>R_{i+1}/\sin\alpha$ never intersect the $i$-th shell, therefore fewer ALPs have the chance to cross the innermost shells. In the right panel of Fig.~\ref{fig:discr} we show the percentage relative error 
\begin{equation}
    \eta = \left(\frac{\left \langle \varepsilon_{\rm dep} \rangle \right\rvert_{N_\alpha}}{\langle \varepsilon_{\rm dep} \rangle \rvert_{N_\alpha\rightarrow \infty}}-1\right) \times 100
\label{eq:eta}
\end{equation}
as a function of the stellar radius $R$, starting from the central radius of the first shell $R^{c}_1=(R_1+R_2)/2\approx 140$~km. It is apparent that for $N_\alpha=2$ spatial energy distribution is not properly reproduced, while the $N_\alpha=10$ approximation agrees with the continuum case down to $R\approx 10^3$~km, with a maximal discrepancy $\eta\lesssim 2\%$ at centre. As $N_\alpha$ increases, the agreement becomes better and better and for $N_\alpha \gtrsim 30$ the result converges to the continuum case. Finally, we stress that for each value of $N_\alpha$ a straight line connects the two values of $\eta$ at $R^c_1$ and $R^c_2 \approx 270$~km, due to the lack of data between these two radii.
\\
Given these uncertainties, we evaluated the bound with $N_\alpha = 10$, since this approximation reproduces the continuum case down to $\sim 10^3$~km. Indeed, for lower radii convection is dominant and we expect that the discrepancy with the continuum limit ($\eta \lesssim 2\%$) does not have a huge impact on the evaluation of the bound.

\section{Comparison of methods}
\label{app:comp}

In previous works, the energy transport due to the ALP production and decay  has been treated as a diffusive process \cite{Raffelt:1988rx}. However, this assumption requires that the ALP mean-free-path  is small compared to the characteristic length over which the energy transport extends.  
As length of reference, we may use the pressure or the temperature scale heights, namely,  $H_P$ or $H_T$, respectively. Note that these parameters well represents the length scale over which the convective energy transport operates in the core of an HB star. The first is given by:
\begin{equation}
    H_P=\left | \frac{dr}{d\ln P} \right |=P\left | \frac{dr}{dP} \right |=P\frac{r^2}{Gm_r\rho} \,\ ,
\label{eq:B1}
\end{equation}
where $m_r=4\pi\int_{0}^{r}r'^2\rho dr'$ is the mass within the radius $r$. Here, we have used the hydrostatic equilibrium equation: $dP/dr=-Gm_r\rho/r^2$. Then, the temperature scale height is given by:
\begin{equation}
    H_T=\left | \frac{dr}{d\ln T} \right |=\left ( \frac{d\ln P}{d\ln T} \right ) \left | \frac{dr}{d\ln P} \right |=\frac{H_P}{\nabla_T} \,\ ,
\label{eq:B2}
\end{equation}
where 
\begin{equation}
H_P= \left | \frac{dr}{d\ln P} \right | \,\ \,\ ,
\,\ \,\ \nabla_T = \left ( \frac{d\ln T}{d\ln P} \right ) \,\ .
\end{equation}

Within the convective core of a HB stars, as well as in the semiconvective layer above it, the energy transport is (practically) adiabatic. Hence, for a gas of monatomic ions and free electrons, the adiabatic temperature gradient is $\nabla_T\sim 2/5$. More outside, in the radiative region of the core, $\nabla_T$ is even smaller. Therefore, $H_T$ is always larger than $H_P$ [see Eq.~\eqref{eq:B2}], and we may conservatively assume that the diffusion approximation is valid if $\lambda_a/H_P\ll 1$. By means of 
Eq.~\eqref{eq:B1}, we find that within the convective core $H_P$ is of the order of $10^5$ km, and smaller outside it.    

More in general, the energy flux at radius $r$ is given by the sum of all contributions to the energy transport, that is:
\begin{equation}
    L_r=L_{\rm conv}+L_\gamma+L_e+L_a \,\ ,
\end{equation}
where the first term represents the energy transported by ascending convective elements, the second represents the forward photon flux, the third is due to the electron thermal conduction, while the last one is the additional (non-standard) term representing the energy transported by means of $photon \rightarrow ALP \rightarrow photon$  processes, as due to  photon coalescence followed by ALP decay. As usual, the transfer of energy due to electromagnetic radiation and electron conduction is driven by the radial temperature gradient (see Ref.~\cite{KW1990}):
\begin{equation}
    L_\gamma=-\frac{4\pi r^2}{3\,\kappa_\gamma\,\rho}\frac{d(aT^4)}{dr}\,, 
\end{equation}
and
\begin{equation}
    L_e=-\frac{4\pi r^2}{3\,\kappa_e\,\rho}\frac{d(aT^4)}{dr}\,, 
\end{equation}
where $a\,T^4$ is the energy stored in the radiation field ($a=\pi^2/15$ in natural units) and $\kappa_\gamma$ and $\kappa_e$  are the photons and the electrons opacities, respectively. 
Then, if the condition for the validity of the ALP diffusion is fulfilled, also $L_a$ may be formally written in the same way, so that:
\begin{equation}
  L_r = L_{conv} - \frac{4\pi r^2}{3\,\kappa\,\rho}\frac{d(aT^4)}{dr}\,, 
  \label{eq:Ltot}
\end{equation}
where
\begin{equation}
    \kappa^{-1} = \kappa_\gamma^{-1} + \kappa_e^{-1} + \kappa_a^{-1}\,.
    \label{eq:ktot}
\end{equation}
 Here, we have introduced the ALP Rosseland mean opacity, as derived by \cite{Raffelt:1988rx,Raffelt:1990yz}:
\begin{equation}
    (\kappa_a\,\rho)^{-1} = \frac{1}{4\,a\,T^3}\int_{m_a}^{\infty} dE\, \beta_E\, \lambda_E\, \frac{\partial B_E}{\partial T}\,,
\label{eq:kappa}
\end{equation}
where $\beta_E=\sqrt{1-m_a^2/E^2}$ is the ALP velocity, $\lambda_E$ is the ALP decay length given by Eq.~\eqref{eq:mean_free_path} and $B_E$ is the ALP thermal spectrum
\begin{equation}
    B_E = \frac{1}{2\pi^2} \frac{E^2 \sqrt{E^2 - m_a^2}}{e^{E/T}-1}\,\ .
\label{eq:BE}    
\end{equation}
Plugging Eqs.~\eqref{eq:mean_free_path} and \eqref{eq:BE} into Eq.~\eqref{eq:kappa}, $\kappa_a$ is found to be
\begin{equation}
\kappa_{a}^{-1} = 7.638 \times 10^{5}\, \textrm{ g cm}^{-2}\,\rho\, g_{a\gamma}^{-2} \left(\frac{m_a}{T}\right)^{-4} T^{-3} \int_{m_a/T}^{\infty} dx \left(1-\left(\frac{m_a/T}{x}\right)^2\right)^{1.5} x^{5} \frac{e^{x}}{(e^x-1)^2}\,,
\label{eq:kcompl}
\end{equation}
with $x=E_a/T$, $g_{a\gamma}$ in GeV$^{-1}$ and $m_a$ and $T$ in keV.\\
In the case of interest $m_a/T\gg 1$, therefore Eq.~\eqref{eq:kcompl} can be rewritten in the relativistic limit, substituting $E_a=m_a+y T$, with $y = \beta^2\,m/2\,T$, and since $m_a/T \gg \beta$ one has
\begin{equation}
\begin{split}
\kappa_{a}^{-1} = &7.638 \times 10^{5}\, \textrm{ g cm}^{-2}\,\rho\, g_{a\gamma}^{-2} \left(\frac{m_a}{T}\right)^{-4} T^{-3}\times \\
&\int_{0}^{\infty}\, dy\,\beta \frac{m_a}{T} \gamma \left(2 \frac{m_a}{T}\,y\right)\,\left( \frac{m_a}{T}^2+2\,\frac{m_a}{T}y\right)\,e^{-m_a/T+y}\,,
\end{split}
\end{equation}
where $\beta = \sqrt{2y\,\frac{T}{m_a}}$ and $\gamma=E_a/m_a=1+yT/m_a$. By integrating over $y$ we can find an analytical expression for $\kappa_{a}^{-1}$, i.e.
\begin{equation}
\kappa_{a}^{-1} = 1.436\times 10^{5}  \textrm{ g cm}^{-2}\, \rho\, g_{a\gamma}^{-2} \left(\frac{m_a}{T}\right)^{-5/2} T^{-3}\, e^{-(m_a/T)} \left[35+\frac{m_a}{T}\left(15+2 \frac{m_a}{T}\right)\right]\,,
\label{eq:non-rel}
\end{equation}
with $g_{a\gamma}$ in GeV$^{-1}$, $m_a$ and $T$ in keV. The approximation in Eq.~\eqref{eq:non-rel} reproduces Eq.~\eqref{eq:kcompl} within $\sim 1 \%$ for all the masses which we are interested in. 

\begin{figure}[t!]
		\vspace{0.cm}
	\includegraphics[scale=0.6] {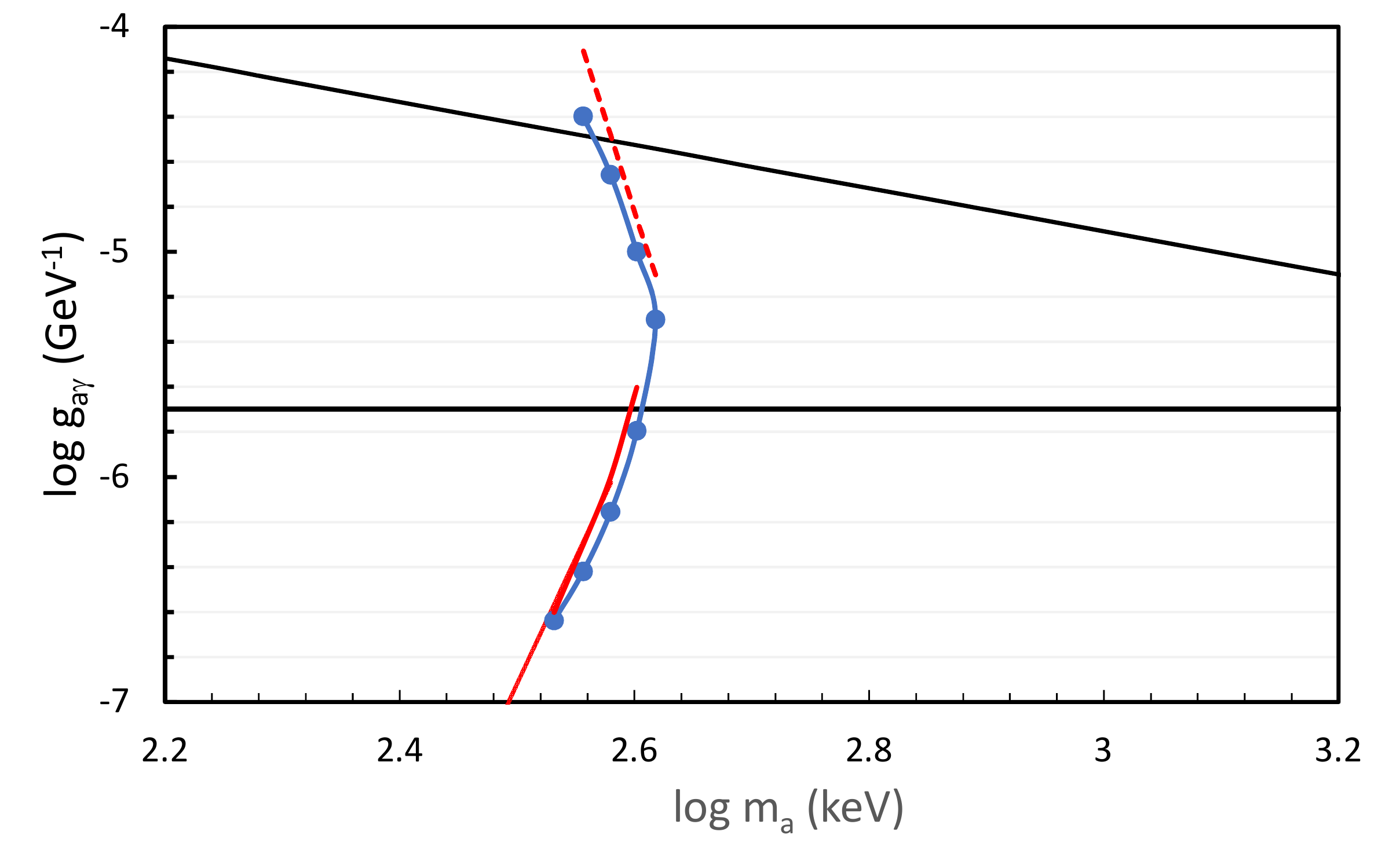}
		\caption{HB bounds in the plane $g_{a\gamma}$ \emph{vs} $m_a$, as computed by adopting: the ballistic model (blue dots and line), the diffusive ALP energy transport (red-dashed line), and the free-streaming approximation (red-solid line). For completeness, the cosmological triangle is also shown.
		}
		\label{fig:comp_bound}
	\end{figure}

In the limit of small ALP mfp, the ballistic model should converge towards the diffusion approximation.  Then, we have modified the FuNS code, by including in the luminosity equation (\ref{eq:Ltot}) the term describing the ALP energy transport. The ALP opacity is computed by means of Eq.~(\ref{eq:non-rel}). In Fig.~\ref{fig:comp_bound} the resulting upper bound is compared to the one obtained with the ballistic model (continuous curve interpolating dots). Only  $m_a$-$g_{a\gamma}$ pairs for which the ALP mean-free-path is smaller than $10^5$ km have been considered (red dashed curve). 
In the opposite limit of large ALP mean-free-path, the ballistic model  reproduces the result obtained by assuming a free streaming (red continuous curve). This picture confirms the expectations and support our new bound for massive ALPs.
\end{document}